# Wireless Full-duplex Medium Access Control for Enhancing Energy Efficiency

Makoto Kobayashi, *Student Member, IEEE*, Ryo Murakami,

Shunsuke Saruwatari, *Member, IEEE* and Takashi Watanabe, *Member, IEEE*

## Abstract

Recent years have witnessed a proliferation of battery-powered mobile devices, e.g., smartphones, tablets, sensors, and laptops, which leads a significant demand for high capacity wireless communication with high energy efficiency. Among technologies to provide the efficiency is full-duplex wireless communication. Full-duplex wireless enhances capacity by simultaneously transmitting uplink and downlink data with limited frequency resources. Previous studies on full-duplex wireless mostly focuses on doubling the network capacity, whereas in this paper we discuss that full-duplex wireless can also provide higher energy efficiency. We propose low power communication by wireless full-duplexing (LPFD), focusing on the fact that the full-duplex communication duration becomes half of the half-duplex communication duration. In the LPFD, by using the sleep state in which the transceiver provided in the wireless communication terminal is turned off, power consumption of the wireless communication terminal is reduced and energy efficiency in wireless full duplex is improved. Simulation results show that the energy efficiency achieved by LPFD is up to approximately 17.3 times higher than the energy efficiency achieved by existing full-duplex medium access protocol. Further, it is up to approximately 27.5 times higher than the energy efficiency using power saving mode of half-duplex communication.

Manuscript received MM DD, YYYY; This work was supported by the JSPS KAKENHI Grant Number JP16H01718. This paper has been presented in part at the IEEE Global Communication Conference (GLOBECOM) Workshop on Full Duplex Wireless Communications, Washington D.C., USA, Dec. 2016. *(Makoto Kobayashi and Ryo Murakami are co-first authors.)*

The authors are with the Department of Information Networking, Graduate School of Information Science and Technology, Osaka University, Suita, Osaka, 565-0871 JAPAN e-mail: {kobayshi.makoto, murakami.ryo, saru, watanabe}@ist.osaka-u.ac.jp.

# I. INTRODUCTION

Wireless full-duplex communication is becoming a reality with the development of interference cancellation [1]–[7]. Wireless full-duplex communication enables simultaneous transmission and reception at the same frequency band. Under current half-duplex communication schemes, transmitter node cannot receive signals from the other node at the same time. This is because a radio wave attenuates sharply over distance, received signal from itself is much greater than received signal from the other node, causes interference. Combining the analog and digital interference cancellation techniques enables wireless full-duplex communication, and hence, double frequency utilization efficiency [1]. Applying wireless full-duplex communication to infrastructure networks [7]–[14] and ad-hoc networks [15]–[18] improves network throughput.

This paper focuses on power consumption of wireless full-duplex communication. Low power consumption of wireless full-duplex communication was investigated in [19]. The full-duplex power saving mode (FDPSM) in [19], which turns media on and off according to a beacon cycle similar to IEEE 802.11 power saving mode (PSM), reduces power consumption. However, the power consumption of FDPSM is just lower than that of full-duplex communication, but higher than the power consumption of existing half-duplex IEEE 802.11 PSM [20]. Additionally, FDPSM cannot achieve high throughput because number of transmission is limited as one in a beacon cycle.

In this paper, we improve the FDPSM by focusing on the fact that wireless full-duplex communication is able to reduce the power consumption of wireless communication. Wireless full-duplex communication makes it possible to combine uplink and downlink communication into the same duration. The occupied time of the full-duplex communication frequency band can be reduced to half the occupied time of wireless half-duplex communication. Additionally, it is possible to share circuits that consume power by simultaneously performing uplink and downlink communication. We introduce the power consumption model in wireless full-duplex communication in Section II.

From the perspective of power consumption, we propose low power wireless communication with full-duplexing and control packets (LPFD-PKT) and low power wireless communication with full-duplex and frequency bitmap (LPFD-FBM). LPFD-PKT reduces power consumption by scheduling bi-directional full-duplex, two-directional full-duplex, and half-duplex communication using buffer information and inter-user interference information at an access point and each user

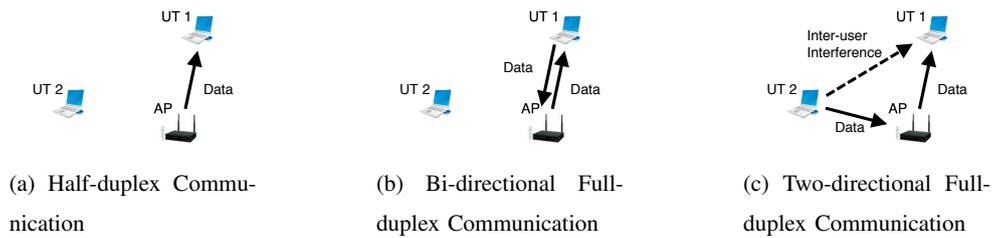

(a) Half-duplex Communication     (b) Bi-directional Full-duplex Communication     (c) Two-directional Full-duplex Communication

Fig. 1. Wireless Network

terminal. LPFD-FBM reduces the packet exchange overhead of LPFD-PKT by using a frequency bitmap. In addition, scheduling scheme in LPFD-PKT and LPFD-FBM enables user terminals to communicate multiple packets in a beacon cycle. In the simulation results, LPFD-FBM achieved up to approximately 9.0 times higher energy efficiency than existing full-duplex communication, and up to approximately 7.0 times higher energy efficiency than IEEE 802.11 PSM.

Methods for exchanging information using orthogonal frequency divisional multiplexing (OFDM) subcarriers have been proposed [21]–[25]. LPFD-FBM uses a frequency bitmap. The frequency bitmap used in LPFD-FBM is an extension of Back2F [21]. Back2F used subcarriers of orthogonal frequency divisional multiplexing (OFDM) to reduce the random backoff time by using an exchange with other terminals. Because the access point and terminal in Back2F are assumed to move the transmitting and receiving circuit simultaneously, it is possible to instantaneously exchange the terminal that has the shortest random backoff. Back2F represents the period of random backoff with subcarriers. In contrast, the frequency bitmap in this paper is used for various purposes, such as collection of buffer information, measurement and collection of interference between terminals, confirmation response, and so on.

The remainder of this paper is organized as follows: Section II introduces the system model in this paper; Section III describes LPFD-PKT, which exchanges information via packets; Section IV presents LPFD-FDM, which reduces the information exchange overhead of LPFD-PKT. Section V discusses the evaluation of the proposed methods. Finally, Section VI concludes the paper.

## II. SYSTEM MODEL

### A. Wireless Full-duplex

This paper assumes a star topology wireless network consisting of one access point (AP) and $N$ user terminals (UTs) equipped with a wireless full-duplex function. Fig. 10 illustrates the network assumed in this paper. The access point and each user terminal communicates by wireless half-duplex communication (Fig. 1a), wireless bi-directional full-duplex communication (Fig. 1b), and wireless two-directional full-duplex communication (Fig. 1c).

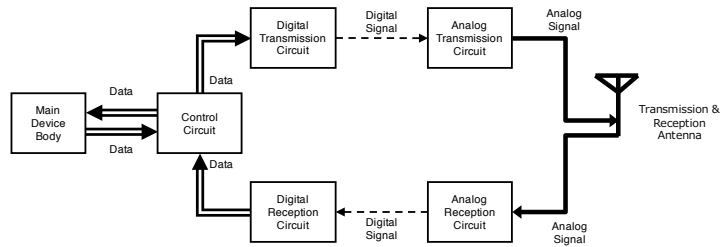

Fig. 2. Model of Half-duplex Communication Circuitss

In wireless half-duplex communication, shown in Fig. 1a, the access point transmits one frame to the user terminal, or the user terminal transmits one frame to the access point in a communication period. Wireless half-duplex communication is currently used in normal wireless local area networks (LANs).

In wireless bi-directional full-duplex communication, shown in Fig. 1b, the access point and user terminal transmit data to each other simultaneously. When performing bi-directional full-duplex communication, both the access point and user terminal must have a frame for each other.

In wireless two-directional full-duplex communication, shown in Fig. 1c, the access point and two user terminals exchange two frames. One user terminal transmits a frame, and the other user terminal receives a frame. The transmitting user terminal must have a frame for the access point. The access point must have a frame for the receiving user terminal. In wireless two-directional full-duplex communication, an inter-user interference problem occurs. The inter-user interference problem is when the transmitting user terminal interferes with the receiving user terminal, which then cannot receive the frame from the access point.

### B. Transceiver Circuits of Wireless Half-duplex Communication

Fig. 2 shows a model of the communication circuit of an existing wireless half-duplex communication terminal. A wireless half-duplex transceiver circuit consists of a main device body, control circuit, digital transmission circuit, analog transmission circuit, digital reception circuit, analog reception circuit, and single transmission/reception antenna.

The main device body represents a personal computer or smart phone, which exchanges packets with the control circuit. The control circuit handles the medium access control (MAC) protocol and ON/OFF switching of the other circuits. The control circuit consumes $P_{\text{CONTROL,ON}}$ [mW] when it is ON, and consumes $P_{\text{CONTROL,OFF}}$ [mW] when it is OFF.

TABLE I

RELATIONSHIP BETWEEN THE WIRELESS HALF-DUPLEX COMMUNICATION TERMINAL STATE AND CIRCUITS

| Communication State | Control Circuit | Transmission Circuits | Reception Circuits |
|---|---|---|---|
| Sleep | OFF | OFF | OFF |
| Transmission (tx) | ON | ON | OFF |
| Reception (rx) | ON | OFF | ON |

The digital transmission circuit performs modulation. The analog transmission circuit converts a digital signal into an analog signal using a DA converter, amplifies the analog signal, and then transmits a radio wave via antenna. In this paper, the combination of the digital transmission circuit and analog transmission circuit is known as transmission circuits. The transmission circuits consume $P_{\text{TX,ON}}$ [mW] when they are ON, and consume $P_{\text{TX,OFF}}$ [mW] when they are OFF.

The digital reception circuit performs demodulation. The analog reception circuit amplifies a received signal, and converts an analog signal into a digital signal by an AD converter. In this paper, the combination of the digital reception circuit and analog reception circuit is known as reception circuits. The reception circuits consume $P_{\text{RX,ON}}$ [mW] when they are ON, and consume $P_{\text{RX,OFF}}$ [mW] when they are OFF.

Table I shows the relationship between the states of wireless half-duplex communication terminal and circuits. In the wireless half-duplex communication power consumption model, the control circuit, transmission circuit, and receive circuit are OFF when the wireless half-duplex communication terminal sleeps; the control circuit and transmission circuit is ON when the wireless half-duplex communication terminal transmits data; and the control circuit and reception circuit is ON when the wireless half-duplex communication terminal receives data.

We can derive the power consumption of the whole transceiver circuit from the relationship between the states of the wireless half-duplex communication terminal and circuits. The power consumption of the whole transceiver circuit of each state ($f_{\text{HD}}(s)|s \in \{\text{sleep, tx, rx}\}$)are

$$f_{\text{HD}}(\text{sleep}) = P_{\text{CONTROL,OFF}} + P_{\text{TX,OFF}} + P_{\text{RX,OFF}}$$

$$f_{\text{HD}}(\text{tx}) = P_{\text{CONTROL,ON}} + P_{\text{TX,ON}} + P_{\text{RX,OFF}}$$

$$f_{\text{HD}}(\text{sleep}) = P_{\text{CONTROL,ON}} + P_{\text{TX,OFF}} + P_{\text{RX,ON}}.$$

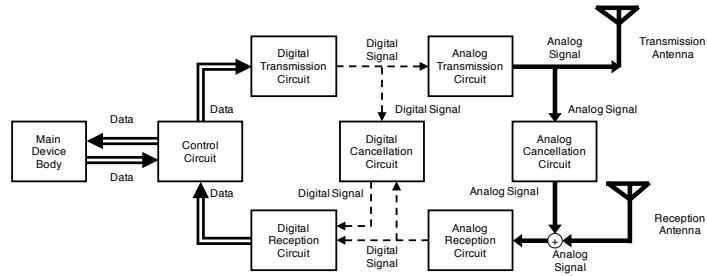

Fig. 3. Model of Full-duplex Communication Circuits

TABLE II

RELATIONSHIP BETWEEN THE STATE OF THE WIRELESS FULL-DUPLEX COMMUNICATION TERMINAL AND CIRCUITS

| Communication State | Control Circuit | Transmission Circuits | Reception Circuits | Cancel Circuits |
|---|---|---|---|---|
| Sleep | OFF | OFF | OFF | OFF |
| Transmission (tx) | ON | ON | OFF | OFF |
| Reception (rx) | ON | OFF | ON | OFF |
| Full-duplex (fd) | ON | ON | ON | ON |

### C. Transceiver Circuits of Wireless Full-duplex Communication

Fig. 3 shows a model of the communication circuit of a wireless full-duplex communication terminal. Wireless full-duplex communication terminals have a digital cancellation circuit and analog cancellation circuit not present in the half-duplex communication terminals. The digital and analog cancellation circuit cancel the self-interference. In this paper, the combination of the digital cancellation circuit and analog cancellation circuit are known as cancellation circuits. The cancellation circuits consume $P_{\text{CANCEL,ON}}$ [mW] when they are ON, and consume $P_{\text{CANCEL,OFF}}$ [mW] when they are OFF.

Fig. 3 shows an example of a full-duplex communication terminal, which has two antennas for transmission and reception, respectively. However, there also exists a full-duplex communication terminal with only one antenna. For example, in [1], the transmission and reception circuit share the antenna by using a circulator for full-duplex communication.

Table II shows the relationship between the states of the wireless full-duplex communication terminal and circuits. The difference from wireless half-duplex communication is that a cancel circuit is added to the circuits and full-duplex communication state is added to the states. The cancel circuits are ON only when the wireless full-duplex communication terminal is in the full-duplex communication state. Additionally, all circuits in the terminal, i.e., the control circuit, transmission circuits, reception circuits, and cancellation circuits, are ON when the state is in full-duplex communication.

We can derive the power consumption of the whole transceiver circuit from the relationship between the states of the wireless full-duplex communication terminal and circuits. The power consumption of the whole transceiver circuit of each state ($f_{\mathrm{HD}}(s)|s \in \{\mathrm{sleep},\ \mathrm{tx},\ \mathrm{rx},\ \mathrm{fd}\}$)are

$$f_{\mathrm{FD}}(\mathrm{sleep}) = P_{\mathrm{CONTROL,OFF}} + P_{\mathrm{TX,OFF}} + P_{\mathrm{RX,OFF}} + P_{\mathrm{CANCEL,OFF}}$$

$$f_{\mathrm{FD}}(\mathrm{tx}) = P_{\mathrm{CONTROL,ON}} + P_{\mathrm{TX,ON}} + P_{\mathrm{RX,OFF}} + P_{\mathrm{CANCEL,OFF}}$$

$$f_{\mathrm{FD}}(\mathrm{sleep}) = P_{\mathrm{CONTROL,ON}} + P_{\mathrm{TX,OFF}} + P_{\mathrm{RX,ON}} + P_{\mathrm{CANCEL,OFF}}$$

$$f_{\mathrm{FD}}(\mathrm{fd}) = P_{\mathrm{CONTROL,ON}} + P_{\mathrm{TX,ON}} + P_{\mathrm{RX,ON}} + P_{\mathrm{CANCEL,ON}}.$$

### D. Bit-per-Joule

This paper uses bit-per-joule (BPJ) [bits/J] as an index of energy consumption efficiency in frame transmission / reception. BPJ is the amount of transmission data per energy. The larger the BPJ, the larger the amount of data that can be transmitted with the same energy consumption. The value of BPJ is the same as the bits per second per watt [bps/W].

In the power consumption model of the wireless full-duplex communication shown in Section 2.3, the average power consumption of terminals is $\bar{P} = \frac{\sum_{s \in S} f_{\mathrm{FD}}(s) t_s}{\sum_{s \in S} t_s}$, where $S(= \{\mathrm{SLEEP},\ \mathrm{TX},\ \mathrm{RX},\ \mathrm{FD}\})$ denotes all states, and $t_s$ represents the duration in the state $s$. The average of throughput is $\bar{R} = \frac{C t_{\mathrm{TX}} + C t_{\mathrm{RX}} + 2 C t_{\mathrm{FD}}}{\sum_{s \in S} t_s} = \frac{C(t_{\mathrm{TX}} + t_{\mathrm{RX}} + 2 t_{\mathrm{FD}})}{\sum_{s \in S} t_s}$. Thus, we can express the value of BPJ [bits/J] as,

$$\mathrm{BPJ} = \frac{\bar{R}}{\bar{P}} = \frac{C(t_{\mathrm{TX}} + t_{\mathrm{RX}} + 2 t_{\mathrm{FD}})}{\sum_{s \in S} f_{\mathrm{FD}}(s) t_s}.$$

## III. Low Power Wireless Communication with Full-duplexing and Control Packets

Based on the power model of Section II, we designed a low power wireless communication with full-duplexing and control packets (LPFD-PKT) that realizes low power wireless communication using wireless full-duplex communication. In LPFD-PKT, the access point assigns each frame to bi-directional full-duplex, two-directional full-duplex, or half-duplex communication. The access point uses information about the buffered packets in the access point, user terminals, and inter-user interference to assign the communication schedule.

### A. Energy Efficiency of Communication Method

The LPFD-PKT design focuses on the fact that the bi-directional full-duplex communication achieves the highest energy efficiency in all communication methods, i.e., bi-directional full-duplex, two-directional full-duplex, and half-duplex communication. $\mathrm{BPJ_{BFD}}$, $\mathrm{BPJ_{TFD}}$, and $\mathrm{BPJ_{HD}}$ is defined as the BPJ of bi-directional full-duplex, two-directional full-duplex, and half-duplex communication, respectively.

First, in bi-directional full-duplex communication, the access point and user terminal transmits $C\tau$ [bits] data, respectively, in the same duration $\tau$. During the transmission duration $\tau$, the user terminal consumes the power for full-duplex communication. Therefore, the BPJ of bi-directional full-duplex communication is shown as,

$$\mathrm{BPJ_{BFD}} = \frac{2 \times C\tau}{f_{\mathrm{FD}}(\mathrm{fd})\tau} = \frac{2C}{P_{\mathrm{CONTROL,\ ON}} + P_{\mathrm{TX,\ ON}} + P_{\mathrm{RX,\ ON}} + P_{\mathrm{CANCEL,\ ON}}}. \tag{1}$$

Moreover, in the two-directional full-duplex communication, the access point and user terminal transmits $C\tau$ [bits] data, respectively, in the same duration $\tau$. Note that two user terminals participate in two-directional full-duplex communication. During the transmission duration $\tau$, one user terminal consumes power for transmission and the other consumes power for reception. The BPJ of two-directional full-duplex communication is

$$\mathrm{BPJ_{TFD}} = \frac{C\tau + C\tau}{f_{\mathrm{FD}}(\mathrm{tx})\tau + f_{\mathrm{FD}}(\mathrm{rx})\tau} = \frac{2C}{2P_{\mathrm{CONTROL,\ ON}} + P_{\mathrm{TX,\ ON}} + P_{\mathrm{RX,\ ON}}}. \tag{2}$$

In half-duplex communication, the access point and user terminal transmit respective $C\tau$ [bits] data time divisionally. The duration of half-duplex communication becomes $2\tau$. The user terminal consumes power for transmission and reception in the respective duration $\tau$. The BPJ of half-duplex communication is

$$\mathrm{BPJ_{HD}} = \frac{C \times 2\tau}{f_{\mathrm{fd}}(\mathrm{tx})\tau + f_{\mathrm{fd}}(\mathrm{rx})\tau} = \frac{2C}{2P_{\mathrm{CONTROL,\ ON}} + P_{\mathrm{TX,\ ON}} + P_{\mathrm{RX,\ ON}}}. \tag{3}$$

By combining (2) and (3), we can obtain $\mathrm{BPJ_{TFD}} = \mathrm{BPJ_{HD}}$. Additionally, we can assume $P_{\mathrm{CONTROL,\ ON}} > P_{\mathrm{CANCEL,\ ON}}$ in general. By combining (1) and (2), and $P_{\mathrm{CONTROL,\ ON}} > P_{\mathrm{CANCEL,\ ON}}$, we can obtain

$$\mathrm{BPJ_{BFD}} > \mathrm{BPJ_{TFD}} = \mathrm{BPJ_{HD}}. \tag{4}$$

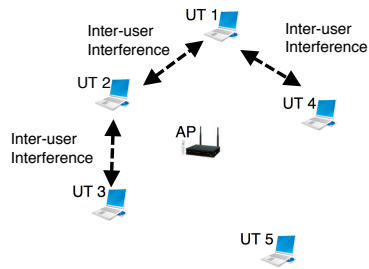

Fig. 4.  Wireless Network: One Access Point and 5 User Terminals

### B. Operation of LPFD-PKT

LPFD-PKT has 4 phases as follows:

1) Beacon frame exchange phase

2) Buffered packet notification and inter-user interference measurement phase

3) Transmission schedule assignment phase

4) Data transmission and acknowledgment phase.

Each user terminal exchanges authentication information when connecting to the access point, and, at the same time, receives the number of user terminals connected to the access point ($N$) and node ID (Node-ID). Node-ID is an integer number from 0 to $N$ and is used for uniquely identifying the access point and user terminal in the network. Node-ID 0 denotes the access point.

Fig. 4 shows an example of wireless network consists of one access point and 5 user terminals. Fig. 5 shows an example of the frame sequence of the LPFD-PKT in the topology composed of the access point and five user terminals shown in Fig. 4. In Fig. 5 and Fig. 4, we assume that the access point has a packet to user terminal 1, and a packet to user terminal 4; user terminal 1 has a packet to the access point; user terminal 3 has two packets to the access point when the beacon frame is exchanged.

The "beacon frame exchange phase" occurs periodically every $t_{\text{beacon}}$, where $t_{\text{beacon}}$ is beacon interval length. The access point periodically transmits the beacon frame. User terminals who receive a beacon frame synchronize their clocks with the access point. The beacon frame indicates the number of user terminals connected to the access point ($N$). In the "buffered packet notification and inter-user interference measurement phase," each user terminal transmits a buffer information (BI) frame and sends the number of buffered packets to the access point. At the same time, user terminals measure inter-user interference. Details regarding the "buffered packet notification and inter-user interference measurement phase" are described in Section III-C. In the "transmission schedule assignment phase," the access point assigns each frame to bi-directional full-duplex, two-directional full-duplex, and half-duplex communication. In order to assign the communication, the access point uses a user interference information request (UIR) frame and user interference information (UII) frame. More details regarding the "transmission schedule assignment phase" are shown in Section III-D. In the "data transmission and acknowledged phase", the access point sends the schedule of transmission by sending the schedule (SCHED) frame. User terminals follow the schedule to transmit and receive data frames. After reception, the access point and user terminals exchange an acknowledgment (ACK) frame. Further detail on the "data transmission and acknowledged phase" is described in Section III-E.

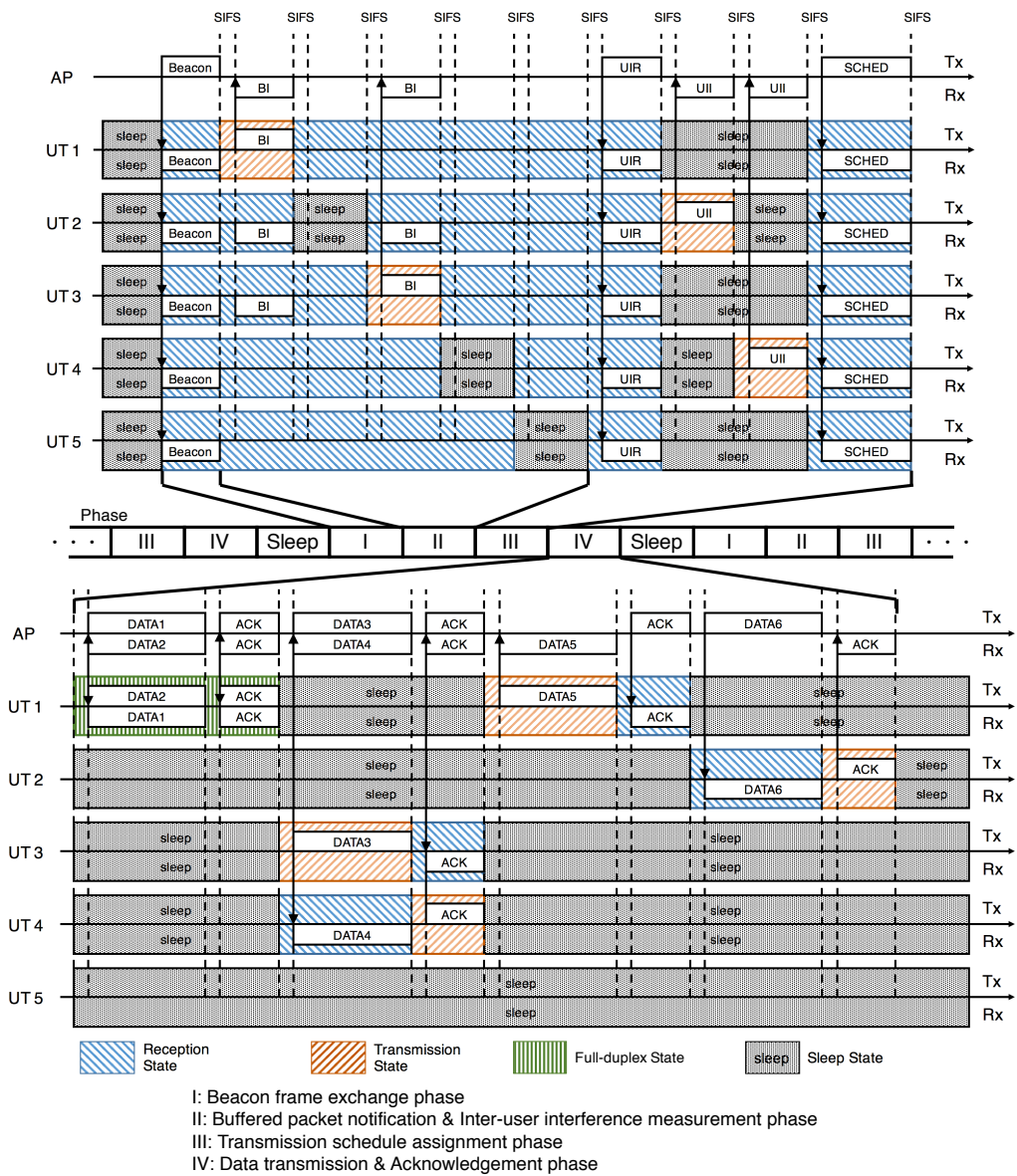

Fig. 5.  Low Power Wireless Communication with Full-duplexing and Control Packets (LPFD-PKT)

### C. Buffered Packet Notification and Inter-user Interference Measurement Phase

User terminals send notification of their buffered packets and measure inter-user interference in order to obtain information to be used in the transmission schedule assignment phase, described in section III-D. Because bi-directional full-duplex communication can enhance the energy efficiency of user terminals, the access point schedules a transmission to create bi-directional full-duplex communication by obtaining the buffered packet in user terminals. Additionally, the access point schedules two-directional full-duplex communication to avoid inter-user interference by gathering information regarding inter-user interference. Inter-user interference wastes the power consumed by the user terminal for downlink reception from the access point, when the access point and user terminals communicate using two-directional full-duplex.

Every user terminal shifts from the sleep state to the awake state at each beacon interval and receives the beacon frame from the access point. Each user terminal transmits the BI frame in the order assigned to the Node-ID after reception of the beacon frame. The BI frame includes the Node-ID of the BI frame source and number of data frames held by the user terminal in the buffer. In the Fig. 5 example, it is assumed that user terminal 1 notifies that it has two frames for the access point, and user terminal 3 notifies that it has one frame for the access point, respectively. In contrast to user terminals 1 and 3, user terminals 2, 4, and 5 shift to the sleep state during the duration of their BI frame transmission because they do not have buffered data for the access point.

All user terminals remain in the awake state for the duration of the BI frame transmission from other user terminals in order to receive the BI frame from these terminals. In the example shown in Fig. 5, user terminal 2 receives BI frames from user terminals 1 and 3. Then, user terminal 2 determines that inter-user interference has occurred between itself and user terminals 1 and 3. User terminal 4 determines that inter-user interference occurred between itself and user terminal 1, because it received the BI frame from user terminal 1. User terminals 1, 3, and 5 determine that they were not influenced by any inter-user interference because they did not receive a BI frame.

### D. Transmission Schedule Assignment Phase

In the "transmission schedule assignment phase," the access point schedules transmission using the received BI frames. The access point assigns the transmission in the order of

  1) bi-directional full-duplex communication

  2) two-directional full-duplex communication

  3) half-duplex communication,

to enhance the energy efficiency keeping the high throughput in response to the result shown in (Fig. 5).

First, the access point assigns bi-directional communication. The access point detects the uplink and downlink pair, which are transmitted by bi-directional communication. In the Fig. 5 example, the access point has a data frame for user terminal 1, and user terminal 1 has two data frames for the access point. Thus, the access point assigns bi-directional full-duplex communication between the access point and user terminal 1 for the first cycle.

Second, the access point assigns two-directional full-duplex communication. The access point collects information regarding inter-user interference via UIR and UII frames in order to determine the uplink and downlink combination of two-directional full-duplex communication. The UIR frame contains the Node-IDs of user terminals that are expected to be participants in two-directional full-duplex communication. Node-IDs in the UIR frame are arranged in the order of UII frame transmission. A user terminal whose Node-ID is included in the UIR frame returns the UII frame to the access point according to the Node-ID order in the UII frame. The UII frame includes the Node-IDs of all user terminals where inter-user interference occurs with the terminal that transmitted the UII frame. Inter-user interference is measured by the BI frame mentioned in Section III-C.

In the Fig. 5 example, the access point requests user terminals 2 and 4, which are candidates for two-directional full-duplex communication, to transmit the UII frame using the UIR frame. User terminals 2 and 4 transmit the UII frame after receiving the UIR frame. User terminal 2 notifies the access point that inter-user interference occurred with user terminals 1 and 3, and user terminal 3 notifies the access point that inter-user interference occurred with user terminal 1.

The access point determines the uplink and downlink combination of two-directional full-duplex communication based on the inter-user interference information. At that time, the access point generates the uplink and downlink combination of two-directional full-duplex communication, such that inter-user interference does not occur. In the Fig. 5 example, the access point knows that user terminal 1 and 3 both have one data frame to transmit. Additionally, the access point buffers one data frame for user terminal 2 and another frame for user terminal 4. The access point assigns a two-directional full-duplex communication (uplink source is user terminal 3; downlink destination is user terminal 4) to the second transmission cycle.

Finally, the access point schedules half-duplex communication. In the duration of the half-duplex communication, the remaining data frames that are not subject to bi-directional or two-directional full-duplex communication are transmitted. In the example shown in Fig. 5, during the third transmission cycle, user terminal 1 transmits a data frame to the access point; and during the fourth transmission cycle, the access point transmits data to user terminal 2.

### E. Data Transmission and Acknowledgment Phase

In the data transmission and acknowledgment phase, first, all user terminals shift from the sleep state to the awake state in order to receive the SCHED frame from the access point. The access point transmits the SCHED frame and notifies all user terminals of the data frame transmission order. The SCHED frame contains information regarding the cycle in which each user terminal will transmit the uplink and downlink. The access point and user terminal transmits and receives the data frame and acknowledgment (ACK) frame according to the schedule described in the SCHED frame. The user terminal sleeps when it does not transmit or receive data, or ACK frames, in order to reduce energy consumption.

In the example shown in Fig. 5, the first cycle of communication is bi-directional full-duplex communication between the access point and user terminal 1. The second cycle communication is two-directional full-duplex communication in which user terminal 3 transmits data to the access point and the access point transmits data to user terminal 4. In the third communication cycle, half-duplex communication from user terminal 1 is performed. Finally, in the fourth communication cycle, half-duplex communication to user terminal 2 is performed.

## IV. Low Power Wireless Communication with Full-duplexing and Frequency Bitmap

In the LPFD-PKT shown in Section III, gathering the information of buffered packets, inter-user interference, and ACK frames, becomes overhead. Particularly, the number of frames for gathering the inter-user interference information increases exponentially with the number of user terminals. Low power wireless communication with full-duplexing and frequency bitmap (LPFD-FBM) also has 4 phases same as LPFD-PKT, e.g., beacon frame exchange phase, buffered packet notification and inter-user interference measurement phase, transmission schedule assignment phase, data transmission and acknowledgment phase. However, in contrast to LPFD-PKT, LPFD-FBM reduces overhead by using a combination of full-duplex communication and frequency bitmap. Specifically, LPFD-FBM uses 5 types of frequency bitmap, namely: buffer information frequency bitmap (BI-FBM), reducing the overhead of gathering the buffered packet information; user interference information request frequency bitmap (UIR-FBM) and user interference information frequency bitmap (UII-FBM), reducing the overhead of gathering the inter-user interference information; schedule frequency bitmap (SCHED-FBM), reducing the overhead of the transmission schedule notification by the access point to user terminals; and acknowledgment frequency bitmap (ACK-FBM), to reduce the acknowledgment overhead.

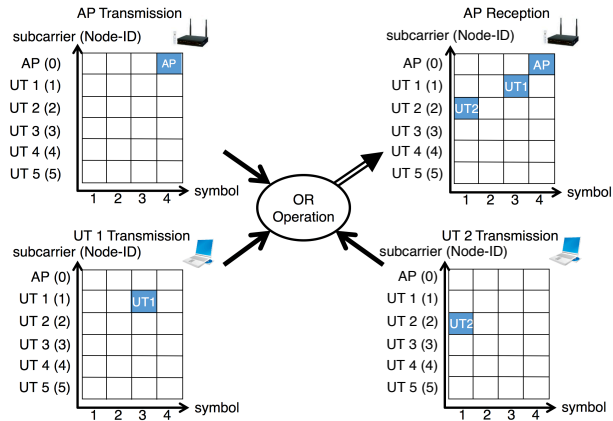

Fig. 6.  Frequency Bitmap

## A. Frequency Bitmap

A frequency bitmap is a technology that allows multiple terminals to simultaneously and bi-directionally transmit information by combining the presence / absence of OFDM signal subcarriers and wireless full-duplex communication. Fig. 6 shows an example of a frequency bitmap. The frequency bitmap has a frequency domain and time domain. The frequency domain of the frequency bitmap is divided by the OFDM signal subcarriers. The time domain of the frequency bitmap is divided by the OFDM symbols. One subcarrier in one OFDM symbol represents one bit. Each user terminal receives the Node-ID, which is assigned to the user terminal when the user terminal associates with the network.

A user terminal or the access point receives duplicated frequency bitmap signals when multiple nodes transmit a frequency bitmap simultaneously. This duplication is similar to an OR operation. By using the frequency bitmap characteristics, user terminals and access points are able to receive information from many nodes simultaneously. Additionally, wireless full-duplex communication enables the ability to perform transmission and reception simultaneously. Therefore, many nodes are able to simultaneously exchange information bi-directionally.

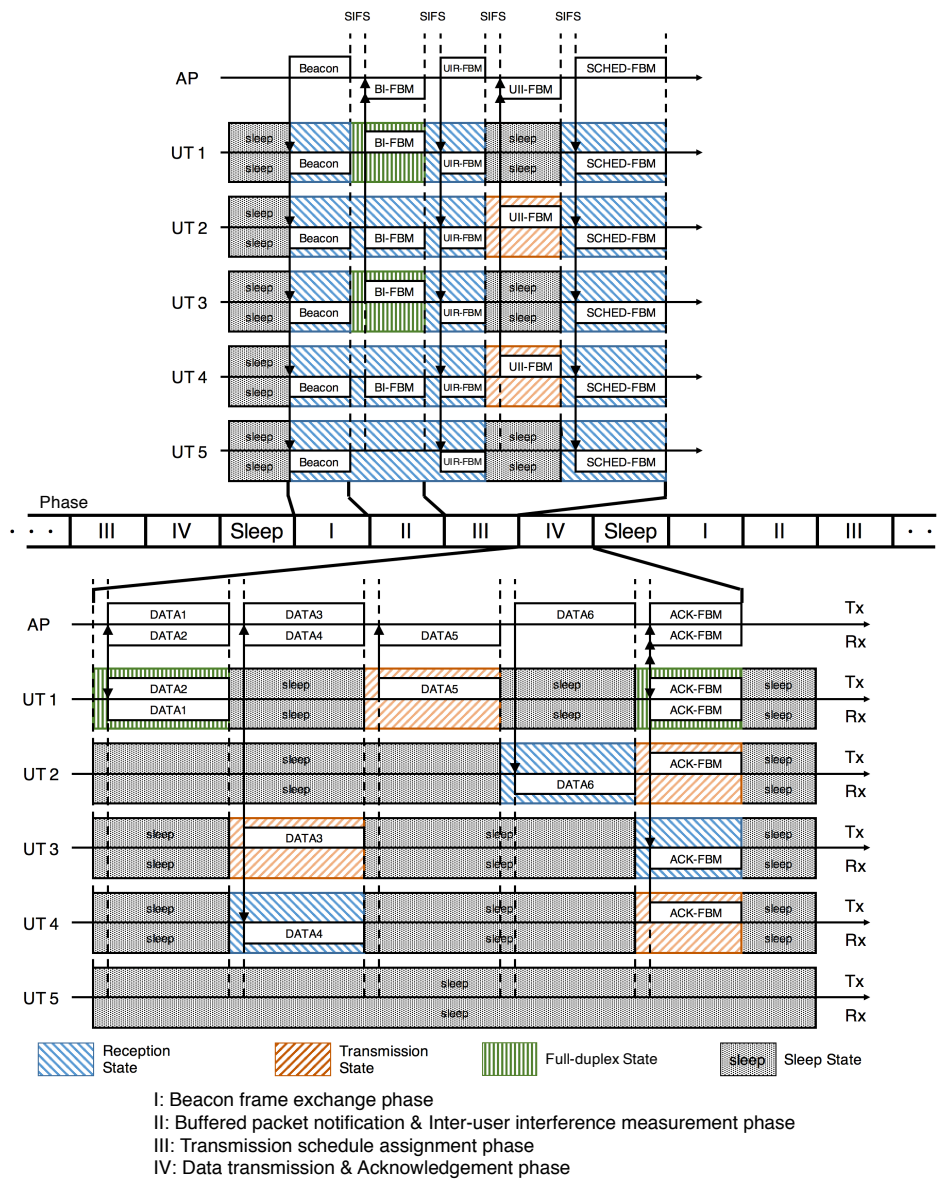

Fig. 7.  Low Power Wireless Communication with Full-duplexing and Frequency Bitmap (LPFD-FBM)

Fig. 7 shows an example of an LPFD-FBM time sequence in a topology composed of one access point and five user terminals, as shown in Fig. 4. In the Fig. 7 example, we assume that the access point has a data frame for user terminal 1 and 4; and user terminal 1 has a data frame for the access point, and user terminal 3 has two data frames at the moment of the beacon frame exchange.

All user terminals shift from the sleep state to the awake state to receive the beacon frame in each beacon interval. LPFD-FBM uses the same beacon frame as the LPFD-PKT frame, mentioned in Section III.

All user terminals who receive the beacon frame, sends a notification of buffered packets in each terminal using BI-FBM. Additionally, at the same time, all user terminals measure inter-user interference by receiving the BI-FBM from other user terminals with full-duplex reception circuits. Details of the BI-FBM will be mentioned in Section IV-B.

The access point schedules bi-directional full-duplex communication after receiving BI-FBMs. Then, the access point informs the reception destination candidates of downlink two-directional full-duplex communication by transmitting a UIR-FBM to all user terminals. User terminals who are mentioned as candidates, transmit the UII-FBM. UII-FBM indicates the buffered packet information and inter-user interference of the transmitting user terminal. More detail regarding the UIR-FBM and UII-FBM is given in Section IV-C and IV-D, respectively.

After receiving the UII-FBM frames, the access point schedules the data frame transmission by using buffered packets in the access point and all user terminals, and inter user interference. The scheduling algorithm uses the same thing as LPFD-PKT. The access point transmits the SCHED-FBM to all user terminals. The SCHED-FBM includes the transmission schedule information. More detail regarding SCHED-FBM is given in Section IV-E.

The user terminal that has received the SCHED-FBM transmits and receives data frames in accordance with the schedule that the access point has determined. After the transmission and reception of all data frames has been completed, the access point and all user terminals exchange the ACK-FBM, which acknowledges data arrival. More detail regarding the ACK-FBM is given in Section IV-F.

### B. Buffer Information Frequency Bitmap (BI-FBM)

The BI-FBM is a frequency bitmap used for buffer status collection used in the buffered packet notification and inter-user interference measurement phase. The BI-FBM replaces the BI frame of LFPD-PKT. Fig. 8a shows an example of BI-FBM. Each sub-carrier is assigned for the access point and user terminals, respectively. Each symbol represents the number of buffered packets in the access point and each user terminal. Each node transmits the symbols of the number of buffered data packets, on the assigned sub-carrier. $L_{\mathrm{MAX}}$ is the maximum number of buffered packets. $L_{\mathrm{MAX}}$ is given a notification from the access point when the user terminal has associated to the access point. Even when a user terminal has more data packets than $L_{\mathrm{MAX}}$ packets, the user terminal transmits $L_{\mathrm{MAX}}$ symbols. In the example shown in Fig. 8a, user terminal 1 buffers two packets, and user terminal 3 has one packet for the access point.

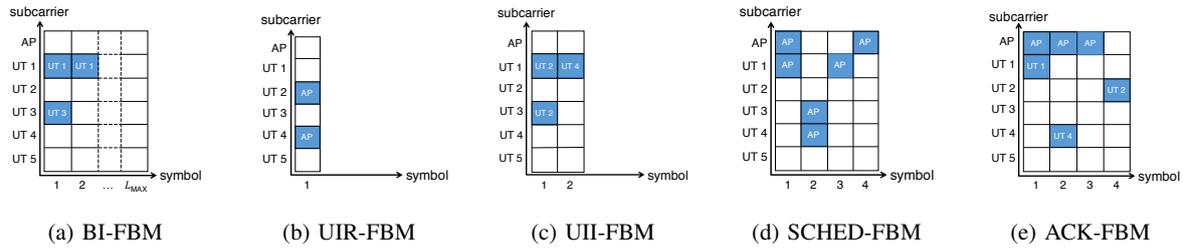

(a) BI-FBM     (b) UIR-FBM     (c) UII-FBM     (d) SCHED-FBM     (e) ACK-FBM

Fig. 8.  Frequency Bitmap used in LPFD-FBM

The BI-FBM is also used for measuring inter-user interference similar to BI frames in LPFD-PKT. When a user terminal receives a BI-FBM containing information that some other Node-ID user terminal has buffered packets, the user terminal determines that inter-user interference occurs with the Node-ID user terminal. In contrast, when a user terminal does not receive any symbols from some Node-ID sub-carriers, the user terminal determines that inter-user interference does not occur with these user terminals.

Even if user terminal A has buffered packets and sends BI-FBM symbols, when user terminal A and B are a hidden terminal relationship pair, the inter-user interference between user terminal A and B does not occur. In that case, the BI-FBM symbol is also attenuated and does not reach user terminal B.

## C. User Interference Information Request Frequency Bitmap (UIR-FBM)

The UIR-FBM is a frequency bitmap that the access point uses to request the inter-user interference information from the user terminal. The UIR-FBM is used in the transmission schedule assignment phase. The UIR-FBM reduces the overhead of the UIR frames in the LPFD-PKT.

Fig. 8b shows an example of the UIR-FBM. The UIR-FBM consists of one symbol. The sub-carrier corresponding to the user terminal, which is a downlink candidate in two-directional full-duplex communication, is set to one by the access point. The access point selects the user terminal based on the buffered packet information gathered by the BI-FBM. In the example shown in Fig. 8b, the sub-carrier for user terminal 2 and 5 is set to one.

*D. User Interference Information Frequency Bitmap (UII-FBM)*

The UII-FBM is a frequency bitmap used for transmitting the inter-user interference between user terminals to the access point from each user terminal. The UII-FBM is used in the transmission schedule assignment phase. The UII-FBM replaces the UII frames of LPFD-PKT and reduces the overhead of UII frames.

Fig. 8c shows an example of UII-FBM. The user terminal that has been requested to send the inter-user interference information by the UIR-FBM from the access point, sends a UII-FBM. The UII-FBM from each user terminal indicates that inter-user interference occurs between the user terminal and two-directional full-duplex destination user terminal mentioned in the UIR-FBM. In order to reduce the UII-FBM size, each user terminal sends inter-user interference information via the following procedure: First, the user terminal receiving the UIR-FBM from the access point, retrieves only the Node-ID which assigned subcarrier is 1, and arranges the retrieved Node-ID ascending order. Note that the retrieved Node-ID represents the user terminal, which is requested to transmit the inter-user interference information from access point in the UIR-FBM. The requested user terminals transmit the UII-FBM in accordance with the order of the retrieved Node-ID. Each user terminal sets the UII-FBM sub-carrier of the Node-ID causing inter-user interference with itself to one.

In the example shown in Fig. 8c, user terminals 2 and 4 send the inter-user interference information. The first symbol is assigned for user terminal 2; and the second is assigned for user terminal 4. User terminal 2 sends a symbol with the sub-carrier for Node-ID 1 and 3 set to one. Also, user terminal 4 sends a symbol with the sub-carrier for Node-ID 1 set to one.

*E. Schedule Frequency Bitmap (SCHED-FBM)*

The SCHED-FBM is a frequency bitmap for the access point to notify all user terminals of the data frame transmission and reception schedule. The SCHED-FBM is used in the transmission schedule assignment phase Fig. 8d shows an example of the SCHED-FBM. Subcarrier denotes the Node-ID of each user terminal and access point. Symbol denotes the scheduled transmission slot. For bi-directional full-duplex communication, the access point transmits a symbol in which two subcarriers for the access point and communication partner user terminal is set to one. In addition, for two-directional full-duplex communication, the access point transmits a symbol in which two subcarriers are set to one, one is a subcarrier for the transmitter user terminal Node-ID, the other is a subcarrier for the receiver Node-ID. For half-duplex communication, the access point transmits a symbol that a subcarrier for a transmitter or receiver user terminal Node-ID is set to one.

In the example shown in Fig. 8d, the SCHED-FBM example shows that bi-directional full-duplex communication between the access point and user terminal 1 will be performed in the first transmission cycle; in the second transmission cycle, two-directional full-duplex communication, an uplink from user terminal 3 and downlink to user terminal 4, will be performed; in the third transmission cycle, an uplink half-duplex communication from user terminal 1 will be performed; and in the fourth transmission cycle, a downlink half-duplex communication to user terminal 2 will be performed.

Note that user terminals are each able to distinguish whether they are transmitter or receiver in two-directional full-duplex communication. This is because only transmission packets or reception packets remain due to bi-directional full-duplex communication being performed just before two-directional full-duplex communication. This mechanism enables the access point to communicate the schedule of two-directional full-duplex communication and half-duplex communication by only using the UIR-FBM and SCHED-FBM.

*F. ACK-FBM*

An example of the ACK-FBM is shown in Fig. 8e. The ACK-FBM is a frequency bitmap used for acknowledgment in the data transmission and acknowledgement phase. Subcarriers represent the Node-IDs of the access point and user terminals; symbols represent transmission cycles. When each access point and user terminal transmits received data in a certain transmission cycle, the access point and user set the symbol representing the transmission cycle to 1 in the ACK-FBM. The access point and all user terminals transmit the ACK-FBMs simultaneously. In the example shown in Fig. 8e, the access point transmits ACK for the uplink transmission from user terminal 1 in the first symbol; the access point transmits ACK for the uplink transmission from user terminal 3 in the second symbol; user terminal 4 transmits ACK for the downlink transmission from user terminal 4 in the third symbol; and user terminal 2 transmits ACK for the downlink transmission from the user terminal 2 in the fourth symbol.

## V. Performance Evaluations

We performed a computer simulation to confirm the performance of LPFD-PKT and LPFD-FBM.

TABLE III

FRAME AND SPACE SIZE

| Frame and Space | Size |
|---|---|
| Data frame | 1528 [byte] |
| ACK frame | 14 [byte] |
| Beacon frame | 28 [byte] |
| PS-Poll frame | 20 [byte] |
| BI frame | 28 [byte] |
| UIR frame | $20 + 6n_{\text{requested}}$ [byte] |
| UII frame | $20 + 6n_{\text{interference}}$ [byte] |
| SCHED frame | $20 + 6n_{\text{scheduled}}$ [byte] |
| One symbol of FBM in LPFD-FBM | 4 [$\mu$sec.] |
| SIFS duration | 16 [$\mu$sec.] |

TABLE IV

POWER CONSUMPTION OF EACH CIRCUIT

| Power Consumption | Value |
|---|---|
| $P_{\text{CONTROL, ON}}$ | $3.00 \times 10^2$ [mW] |
| $P_{\text{TX, ON}}$ | $5.25 \times 10^2$ [mW] |
| $P_{\text{RX, ON}}$ | $1.95 \times 10^2$ [mW] |
| $P_{\text{CONTROL, OFF}}$ | 49.5 [mW] |
| $P_{\text{TX, OFF}}$ | 0.00 [mW] |
| $P_{\text{RX, OFF}}$ | 0.00 [mW] |
| $P_{\text{CANCEL, OFF}}$ | 0.00 [mW] |
| $P_{\text{CANCEL, ON}}$ | $0.00 - 5.00$ [mW] |

## A. Evaluation Environment

The star topology of the evaluation environment contains one access point and $N$ user terminals. User terminals are settled randomly in a sphere of a 5 [m] radius centered on the access point. If the distance between one user terminal and another user terminal is within 5 [m], inter-user interference occurs. Uplink traffic from each user terminal ($\lambda_{\text{up}}$) and downlink traffic for each terminal ($\lambda_{\text{down}}$) arrives according to a Poisson arrival with $\lambda(= \lambda_{\text{up}} = \lambda_{\text{down}})$ [frames/sec.] arrival rate. Frame losses occur when the frame collides with another frame. The access point and user terminals transmit data with a fixed data rate of 6 [Mbps] and data size. Beacon interval length ($t_{\text{beacon}}$) is 100 [ms]. We assume that the access point and user terminals can perform ideal self-interference cancellation. Thus, there is no rate reduction due to self-interference. The maximum number of buffered packet notifications from each user terminal and access point is $L_{\text{max}} = 40$ in the LPFD-FBM.

The size of each frame is defined based on the IEEE 802.11 standard [20]. Table III shows the size of each frame. The size of data frame is defined as 1528 [byte]; The size of ACK frame is defined as 14 [byte]; The beacon frame size is defined as 28 [byte]; The PS-Poll frame size is defined as 20 [byte]. In this paper, the frame sizes used in the LPFD-PKT are defined as follows: the BI frame is 28 [byte]; the UIR frame is $20 + 6n_{\text{requested}}$ [byte] when the access point requests $n_{\text{requested}}$ user terminals to transmit inter-user interference information; the UII frame is defined as $20 + 6n_{\text{interference}}$ [byte] when each user terminal communicates interference with $n_{\text{interference}}$ user terminals; and the SCHED frame is $20 + 6n_{\text{scheduled}}$ [byte] when the access point communicates the schedule of $n_{\text{scheduled}}$ cycles data and ACK transmission. Additionally, the length of one symbol in the frequency bitmap used in the LPFD-FBM is defined as 4 [$\mu$sec.]. SIFS duration is defined as 16 [$\mu$sec.] [20].

Power consumption of each circuit is defined based on the chipset of Wi-Fi, SX-SDCAG 802.11a/b/g SDIO Card Module [26]. Table IV shows the power consumption of each circuit. $P_{\text{CONTROL, ON}}$ is $4.95 \times 10^1$ [mW]; $P_{\text{TX, ON}}$ is $7.76 \times 10^2$ [mW]; $P_{\text{RX, ON}}$ is $4.46 \times 10^2$ [mW]; $P_{\text{CONTROL, OFF}}$ is 2.00 [mW]; $P_{\text{TX, OFF}}$ is 0.00 [mW]; $P_{\text{RX, OFF}}$ is 0.00 [mW]; $P_{\text{CANCEL, OFF}}$ is 0.00 [mW]. Currently, no off-the-shelf self-interference cancellation circuit for full-duplex WLAN exists. The power consumption of cancelations, e.g., balun cancelation [1], [4], passive cancelation [3], is small enough to neglect power consumption. In this paper, we evaluated with variable $P_{\text{CANCEL, ON}}$.

We compared the performance of the following five approaches to benchmark the LPFD-PKT and LPFD-FBM:

1) LPFD-PKT: LPFD-PKT is the proposed method mentioned in Section III.

2) LPHD-PKT: LPHD-PKT is LPFD-PKT without full-duplex communication. The difference between LPHD-PKT and LPFD-PKT is transmission schedule assignment. In transmission schedule assignment phase of LPHD-PKT, the access point assigns the transmission only by half-duplex communication.

3) LPFD-FBM: LPFD-FBM is also the proposed method mentioned in Section IV. LPFD-FBM reduces the overhead of frame exchanges in LPFD-PKT by using a frequency bitmap.

4) Full-duplex continuous active mode (FDAM): FDAM is the normal full-duplex communication mode. MAC of FDAM is designed based on [4]. Nodes in [4] start data transmission with CSMA/CA. Bi-directional full-duplex communication is supported, not but two-directional full-duplex communication in [4]. FDAM has reduced energy consumption compared to the LPFD-PKT and LPFD-FBM.

5) Half-duplex power saving mode (HDPSM): HDPSM is the existing half-duplex power saving mode in IEEE 802.11 [20].

6) Ideal LPFD: Ideal LPFD is the theoretical ideal performance of the low power full-duplex communication shown in Appendix. Ideal LPFD shows the performance of LPFD-PKT and LPFD-FBM when information exchange overhead does not exist. Ideal LPFD represents a theoretical limitation of LPFD-PKT and LPFD-FBM.

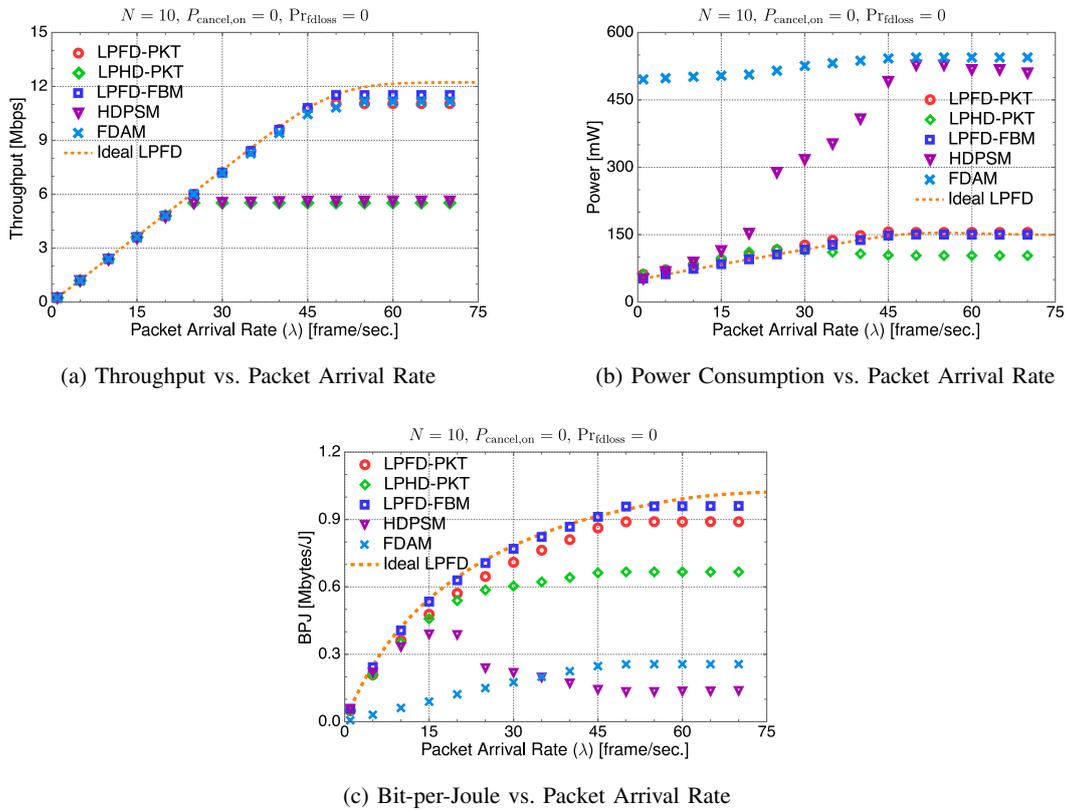

(a) Throughput vs. Packet Arrival Rate

(b) Power Consumption vs. Packet Arrival Rate

(c) Bit-per-Joule vs. Packet Arrival Rate

Fig. 9.  Evaluation for Increasing Packet Arrival Rate

## B. Evaluation of Throughput for Increasing Packet Arrival Rate

First, we evaluated the throughput with an increasing packet arrival rate ($\lambda$). Fig. 9a shows the throughput when the packet arrival rate changes from 1 to 70 [frames/sec.], the number of user terminals ($N$) is fixed at 15, power consumption of cancellation circuit ($P_{\text{CANCEL, ON}}$) is fixed at 0.0 [mW], and complete self-interference cancellation is available ($\text{Pr}_{\text{fdloss}} = 0$).

Fig. 9a shows that the throughput of proposed methods, LPFD-FBM and LPFD-PKT, achieved close to the theoretical limit. This is because the proposed methods efficiently scheduled transmission based on the traffic. However, the difference between throughput of the theoretical limitation and the proposed method appears because of overhead. LPFD-FBM achieved the higher throughput compared to LPFD-PKT, because the frequency bitmap reduces the overhead of LPFD-PKT. Additionally, LPFD-PKT and LPFD-FBM achieved approximately twice the throughput of HDPSM or LPHD-PKT ($\lambda \geq 50$ [frame/sec.]). The reason is that full-duplex communication is able to double the capacity.

### C. Evaluation of Power Consumption for Increasing Packet Arrival Rate

In this section, we evaluated the power consumption of one user terminal with increasing packet arrival rate $\lambda$. Fig. 9b shows the power consumption of a user terminal when the packet arrival rate changes from 1 to 70 [frames/sec.], the number of user terminals ($N$) is fixed at 15, power consumption of cancellation circuit ($P_{\text{CANCEL, ON}}$) is fixed at 0.0 [mW], and complete self-interference cancellation is available ($\text{Pr}_{\text{fdloss}} = 0$).

Fig. 9b shows that power consumption of LPFD-PKT, LPHD-PKT, and LPFD-FBM are three of smallest. This is because the proposed scheduling methods reduce power consumption. Power consumption of LPHD-PKT is smallest when the packet arrival rate is larger than 30 [packets/sec.] because LPHD-PKT treats approximately half number of packets LPFD-PKT and LPFD-PKT treats.

Additionally, both proposed methods, LPFD-PKT and LPFD-FBM, achieved low power consumption close to the theoretical limitation (Ideal LPFD). This is because the transmission scheduling algorithm of the proposed methods efficiently manage energy.

### D. Evaluation of Bit-per-Joule for Increasing Packet Arrival Rate

Section V-B and Section V-C show the performance of throughput and power consumption, respectively. In this section, we evaluate the BPJ with increasing packet arrival rate ($\lambda$) to show the energy efficiency of the proposed methods. Fig. 9c shows the BPJ of a user terminal when the packet arrival rate changes from 1 to 70 [frames/sec.], the number of user terminals ($N$) is fixed at 15, power consumption of the cancellation circuit ($P_{\text{CANCEL, ON}}$) is fixed at 0.0 [mW], and complete self-interference cancellation is available ($\text{Pr}_{\text{fdloss}} = 0$).

Fig. 9c shows that LPFD-FBM achieved the highest BPJ, and LPFD-PKT is second highest. LPFD-FBM achieved approximately 1.4 times the BPJ of HDPSM ($\lambda = 15$[frame/sec.]) and up to approximately 7.0 times the BPJ of HDPSM ($\lambda = 70$[frame/sec.]), approximately 6.0 times the BPJ of FDAM ($\lambda = 15$[frame/sec.]), and approximately 9.0 times the BPJ of FDAM ($\lambda = 1$[frame/sec.]). Additionally, BPJ of both proposed methods, LPFD-PKT and LPFD-FBM, approached the theoretical limitation with an increasing packet arrival rate. This is because control packet transmission is dominant as compared with data frame transmission under a low packet arrival rate, and decreases the BPJ. The LPFD-FBM achieved higher BPJ compared to that of LPFD-PKT. This is because the frequency bitmap reduces the control frame overhead. On the other hand, the BPJ of HDPSM reduced with the increasing packet arrival rate. In the

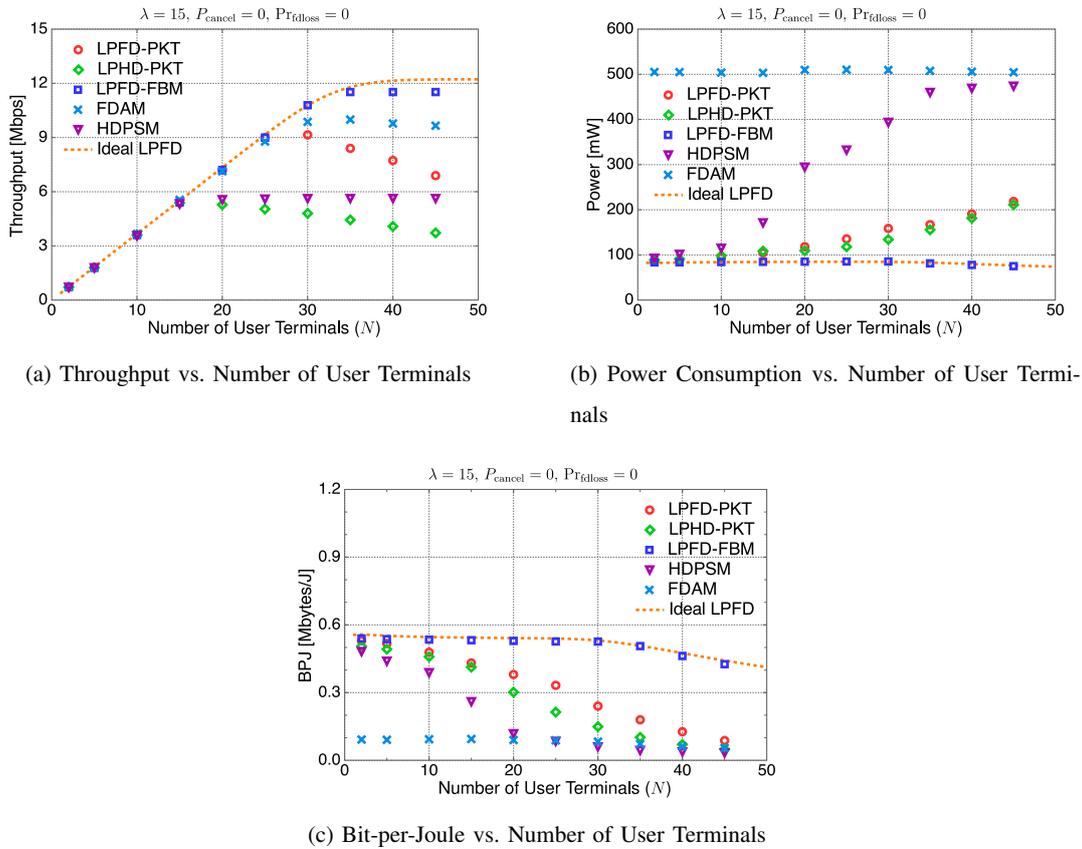

(a) Throughput vs. Number of User Terminals

(b) Power Consumption vs. Number of User Terminals

(c) Bit-per-Joule vs. Number of User Terminals

Fig. 10.   Evaluation for Increasing Number of User Terminals

HDPSM, each user terminal gains the transmission and reception right by contention mechanism. This contention mechanism forces the user terminal to continue awake mode until the end of the transmission in the beacon interval.

### E. Evaluation for Increasing Number of User Terminals

From Section V-B to V-D, the performance evaluation with a fixed number of user terminals is reported. In this section, we evaluate the performance with a variable number of user terminals, where packet arrival rate ($\lambda$) is fixed at 15 [frame/sec.], power consumption of cancellation circuit ($P_{\text{CANCEL, ON}}$) is fixed at 0.0 [mW], and perfect self-interference cancellation is available ($\text{Pr}_{\text{fdloss}} = 0$).

First, we show the throughput evaluation. Fig. 10a shows the throughput when the number of user terminals ($N$) changes from 2 to 45. Fig. 10a shows the throughput of proposed methods, LPFD-FBM and LPFD-PKT, achieved close to the theoretical limitation. This is because the proposed methods efficiently scheduled transmission based on traffic. The LPFD-FBM achieved the highest throughput because the frequency bitmap reduces the LPFD-PKT overhead. The throughput of LPFD-PKT decreases with the increasing number of user terminal where the number of user terminal is larger than 30 ($N \geq 30$); and the throughput of LPHD-PKT decreases with the increasing the number of user terminal where the number of user terminal is larger than 15 ($N \geq 15$). The overhead of information exchange by packets increases with the number of user terminals, and suppress the data transmission. Additionally, LPFD-FBM achieved approximately twice the throughput of the HDPSM where the number of user terminals is larger than 35 ($N \geq 35$). This is because full-duplex communication is able to double the capacity.

Second, we show the power consumption evaluation. Fig. 10b shows the power consumption of one user terminal when the number of user terminals changes from 2 to 45. Fig. 10b shows the power consumption of the LPFD-FBM is the lowest. Both proposed methods, LPFD-PKT and LPFD-FBM, maintained the low power consumption with the increasing number of user terminals. The power consumption of LPFD-PKT and LPFD-FBM are less than approximately one-fifth of the HDPSM. In contrast, the power consumption of the HDPSM increased with the number of user terminals. This is because the contention mechanism used in HDPSM increases overhead with the number of competitors (user terminals).

Last, we show the BPJ evaluation. Fig. 10c shows the BPJ when the number of user terminals changes from 2 to 45. Fig. 10c shows that LPFD-FBM achieved the highest BPJ. The BPJ of the LPFD-FBM is approximately 2.9 times higher than that of LPFD-PKT, and approximately 8.8 times higher than that of the HDPSM when the number of user terminals is 30. This is because the frequency bitmap used in the LPFD-FBM reduces the overhead of exchanging LPFD-PKT buffer information. When the number of user terminals is large, the reduction of overhead owing to the frequency bitmap effect is higher.

### F. Evaluation for Increasing Cancellation Circuit Power Consumption

There is no off-the-shelf self-interference cancellation circuit for full-duplex communication. Power consumption of the self-interference cancellation circuit is expected to be very low. From Section V-B to V-E, the performance evaluation when the power consumption of the cancellation circuit ($P_{\text{CANCEL, ON}}$) is fixed at 0.0 [mW]. However, it is expected that the energy efficiency (BPJ) of the proposed method reduces with the increase of the cancellation circuit power consumption. We evaluated the power consumption of one user terminal and BPJ when the power consumption of the interference cancellation changes from 0.0 [mW] to 5.0 [mW], Where the number of user

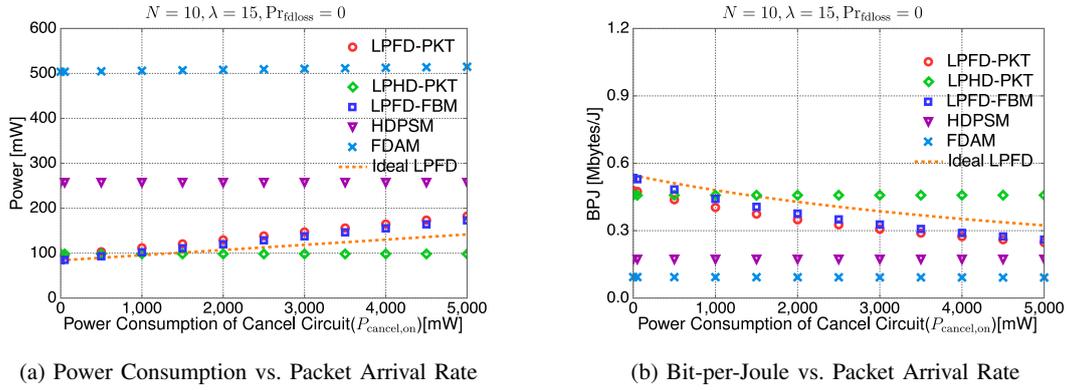

(a) Power Consumption vs. Packet Arrival Rate

(b) Bit-per-Joule vs. Packet Arrival Rate

Fig. 11.  Evaluation for Increasing Cancellation Circuit Power Consumption

terminals ($N$) is fixed at 10, packet arrival rate ($\lambda$) is fixed at 15 [frame/sec.], and perfect self-interference cancellation is available ($\mathrm{Pr}_{\mathrm{fdloss}} = 0$).

First, we show the evaluation of power consumption. Fig. 11a shows the power consumption of one user terminal when the power consumption of interference cancellation changes from 0.0 [mW] to 5.0 [mW]. The power consumption of LPFD-PKT, LPFD-FBM, and FDAM increased linearly with the increasing power consumption of the cancel circuit.

When the power consumption of the cancellation circuit is less than 5.0 [mW], the power consumption of one user terminal of the proposed methods, LPFD-PKT and LPFD-FBM, is less than HDAM.

Additionally, we evaluate the BPJ. Fig. 11b shows the BPJ when the power consumption of the interference cancellation changes from 0.0 [mW] to 5.0 [mW]. The evaluation result of the BPJ also shows the same trend as the power consumption evaluation result. When the power consumption of the cancellation circuits is less than 5.0 [mW], the BPJ of the proposed methods, LPFD-PKT and LPFD-FBM, is less than HDAM.

## G. Evaluation for Imperfect Full-duplex Cancellation

From Section V-B to Section V-F, we assume perfect full-duplex cancellation. In this section, we assume a situation where full-duplex performance drops due to imperfect full-duplex cancellation. In order to evaluate incomplete full-duplex cancellation, we define full-duplex data loss rate ($\mathrm{Pr}_{\mathrm{fdloss}}$). According to the rate $\mathrm{Pr}_{\mathrm{fdloss}}$, the packet transmitted by full-duplex communication is lost because of imperfect cancellation.

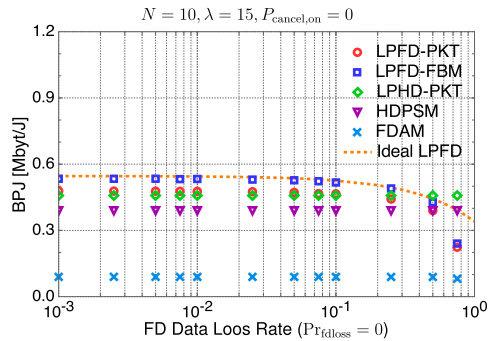

Fig. 12.  Bit-per-Joule vs. Full-duplex Cancelation Performance

Fig. 12 shows the BPJ performance when full-duplex data loss rate ($\text{Pr}_{\text{fdloss}}$) changes from $1.0 \times 10^{-3}$ to $0.75 \times 10^{0}$. The LPFD-FBM achieves the highest BPM when the full-duplex data loss rate is from $1.0 \times 10^{-3}$ to $0.25 \times 10^{0}$. The BPJ of the proposed methods, LPFD-PKT and LPFD-FBM, decrease with the increasing full-duplex data loss rate. However, the proposed methods, LPFD-PKT and LPFD-FBM, achieve higher BPM, compared to HDPSM, even when full-duplex data loss rate ($\text{Pr}_{\text{fdloss}}$) is $0.5 \times 10^{0}$. Additionally, the BPJ of LPFD-FBM is approximately 1.0 [Mbyte/J] , and more than 95% of the highest BPJ of LPFD-FBM when the full-duplex data loss rate is $1.0 \times 10^{-1}$. On the other hand, the BPJ of ideal LPFD suggests that the BPJ of LPFD cannot be higher than HDPSM when the full-duplex data loss rate is larger than $0.7 \times 10^{0}$.

## VI. Conclusion

In this paper, we proposed LPFD-PKT and LPFD-FBM to reduce energy consumption and enhance energy efficiency (Bits-per-Joule) of user terminals in wireless networks. Both LPFD-PKT and LPFD-FBM enhance energy efficiency by scheduling bi-directional full-duplex communication, two-directional full-duplex communication, and half-duplex communication. Additionally, LPFD-FBM reduces the overhead of exchanging buffered packet information in LPFD-PKT by using a frequency bitmap. Performance evaluation shows that both LPFD-PKT and LPFD-FBM achieve higher energy efficiency compared to the IEEE 802.11 PSM and existing full-duplex MAC protocol.

## Appendix

### Ideal Low Power Full-duplex Communication

In ideal LPFD, arrived packets are scheduled for transmission perfectly without any overhead for scheduling. We derive expected values of the number of packets transmitted by bi-directional full-duplex, two-directional full-duplex, and half-duplex communication. Before deriving expected values, we define the probability $p(\lambda, \tau, k)$ of $k$ packet arrivals in a time interval $\tau$ becomes $p(\lambda, \tau, k) = \exp(-\lambda\tau)\frac{(\lambda\tau)^k}{k!}$, where the packet arrival rate is $\lambda$.

Then, we can get the expected number of packets transmitted by full-duplex ($[N_{\text{FD}}]$), and that of half-duplex communication ($E[N_{\text{HD}}]$) as

$$
\begin{aligned}
E[N_{\text{FD}}] = 2 &\sum_{k=1}^{N_{\max}-1} k\, p(N\lambda_{\text{down}}, t_{\text{beacon}}, k) p(N\lambda_{\text{up}}, t_{\text{beacon}}, k) \\
&+ 2 \sum_{k=1}^{N_{\max}-1} k\, p(N\lambda_{\text{down}}, t_{\text{beacon}}, k) \left\{ 1 - \sum_{j=0}^{k} p(N\lambda_{\text{up}}, t_{\text{beacon}}, j) \right\} \\
&+ 2 \sum_{k=1}^{N_{\max}-1} k \left\{ 1 - \sum_{j=0}^{k} p(N\lambda_{\text{down}}, t_{\text{beacon}}, j) \right\} p(N\lambda_{\text{up}}, t_{\text{beacon}}, k) \\
&+ 2 N_{\max} \left( 1 - \sum_{k=0}^{N_{\max}-1} p\left(N\lambda_{\text{down}}, t_{\text{beacon}}, k\right) \right) \left( 1 - \sum_{k=0}^{N_{\max}-1} p\left(N\lambda_{\text{up}}, t_{\text{beacon}}, k\right) \right),
\end{aligned}
\tag{5}
$$

$$
\begin{aligned}
E[N_{\text{HD}}] = &\sum_{k=1}^{\infty} \sum_{j=0}^{\min(k, N_{\max})} (\min(k, N_{\max}) - j) p(N\lambda_{\text{down}}, t_{\text{beacon}}, k) p(N\lambda_{\text{up}}, t_{\text{beacon}}, j) \\
&+ \sum_{k=1}^{\infty} \sum_{j=0}^{\min(k, N_{\max})} (\min(k, N_{\max}) - j) p(N\lambda_{\text{down}}, t_{\text{beacon}}, j) p(N\lambda_{\text{up}}, t_{\text{beacon}}, k)
\end{aligned}
\tag{6}
$$

where $N_{\max}$ is the maximum number of communication pairs (full-duplex or half-duplex) in one beacon cycle, $N_{\max} = \lfloor \frac{\text{Beacon Interval}}{\text{Data frame size / Data rate}} \rfloor (= \lfloor \frac{100[\text{ms}]}{1528[\text{byte}]/6[\text{Mbps}]} \rfloor = 49)$.

Then, we can obtain the expected number of packets transmitted by two-directional full-duplex communication as

$$
\begin{aligned}
E[N_{\text{TFD}}] = 2 &\sum_{k=1}^{N_{\max}-1} f(k, k) p(N\lambda_{\text{down}}, t_{\text{beacon}}, k) p(N\lambda_{\text{up}}, t_{\text{beacon}}, k) \\
&+ 2 \sum_{k=1}^{N_{\max}-1} \sum_{j=k+1}^{\infty} f(k, j) p(N\lambda_{\text{down}}, t_{\text{beacon}}, k) p(N\lambda_{\text{up}}, t_{\text{beacon}}, j) \\
&+ 2 \sum_{k=1}^{N_{\max}-1} \sum_{j=k+1}^{\infty} f(k, j) p(N\lambda_{\text{down}}, t_{\text{beacon}}, j) p(N\lambda_{\text{up}}, t_{\text{beacon}}, k) \\
&+ 2 \sum_{k=N_{\max}}^{\infty} f(N_{\max}, k) p(N\lambda_{\text{down}}, t_{\text{beacon}}, k) p(N\lambda_{\text{up}}, t_{\text{beacon}}, k) \\
&+ 2 \sum_{k=N_{\max}}^{\infty} \sum_{j=k+1}^{\infty} f(N_{\max}, j) p(N\lambda_{\text{down}}, t_{\text{beacon}}, k) p(N\lambda_{\text{up}}, t_{\text{beacon}}, j)
\end{aligned}
$$

$$+ 2 \sum_{k=N_{\max}}^{\infty} \sum_{j=k+1}^{\infty} f(N_{\max}, j) p(N\lambda_{\text{down}}, t_{\text{beacon}}, j) p(N\lambda_{\text{up}}, t_{\text{beacon}}, k). \tag{7}$$

where $f(k, j) = \sum_{i=0}^{k} i \left(\frac{1}{N}\right)^{(k-i)} \left(\frac{N-1}{N}\right)^{\{i+(j-k)\}} \binom{j}{k-i}$. Function $f(k, j)$ represents the bi-directional transmission probability when $k$ full-duplex pairs are communicated, and $j$ is larger number of packets arrived in access point or user terminals. Combining (5), (7), the expected number of packets transmitted by two-directional full-duplex communication is

$$E[N_{\text{BFD}}] = E[N_{\text{FD}}] - E[N_{\text{TFD}}]. \tag{8}$$

Using (1) - (3) and (5) - (8), we can obtain the bit-per-joule of ideal LPFD as

$$\text{BPJ}_{\text{ideal}} = \frac{\text{BPJ}_{\text{BFD}} E[N_{\text{BFD}}] + \text{BPJ}_{\text{TFD}} E[N_{\text{TFD}}] + \text{BPJ}_{\text{HD}} E[N_{\text{HD}}]}{E[N_{\text{BFD}}] + E[N_{\text{TFD}}] + E[N_{\text{HD}}]}.$$